\documentclass[a4paper,twocolumn, nofootinbib,superscriptaddress,aps,pra, 14pt,eqsecnum,notitlepage,showkeys]{revtex4-1}
\usepackage{multirow}
\usepackage{graphicx}
\usepackage{bm}
\usepackage{dcolumn}
\usepackage{hyperref}
\usepackage{gensymb}
\usepackage{amsmath}
\usepackage{dcolumn}    
\usepackage{ifpdf}
\usepackage{amssymb,lineno,amsfonts}
\usepackage{graphicx}   
\usepackage{bm}         
\usepackage{bbm}
\usepackage{mathrsfs}
\usepackage{upgreek}
\usepackage{mathtools}
\usepackage{epstopdf}
\usepackage{setspace}
\usepackage{hyperref}
\usepackage{natbib}
\usepackage{esvect}
\usepackage{amsmath}
\usepackage[usenames,dvipsnames]{xcolor}
\definecolor{med-blue}{RGB}{25,25,112}
\hypersetup{colorlinks, linkcolor={blue},citecolor={blue}, urlcolor={blue}}
\usepackage{times }
\usepackage [english]{babel}
\usepackage [autostyle, english = american]{csquotes}
\MakeOuterQuote{"}
\begin{document}
\title{Identification of a Griffiths singularity in a geometrically frustrated antiferromagnet}
\author{Jitender Kumar}
\author{Soumendra Nath Panja}
\author{Shanu Dengre}
\affiliation{Department of Physics, Indian Institute of Science Education and Research}
\author{Sunil Nair}
\affiliation{Department of Physics, Indian Institute of Science Education and Research}
\affiliation{Centre for Energy Science, Indian Institute of Science Education and Research,\\ Dr. Homi Bhabha Road, Pune, Maharashtra-411008, India}
\date{\today}
\begin{abstract} 
We report the observation of a Griffiths Phase in the geometrically frustrated antiferromagnet DyBaCo${_4}$O${_{7+\delta}}$. Its onset is determined using measurements of the thermoremanent magnetization, which is shown to be superior to conventional in-field measurement protocols for the identification of the Griffiths Phase. Within this phase, the temporal relaxation of magnetization exhibits a functional form which is expected for Heisenberg systems, reflecting the nature of spin interactions in this class of materials. Interestingly, the effective Co${^{2+}}$/Co${^{3+}}$ ratio tailored by varying the oxygen non-stoichiometry $\delta$ is only seen to influence the antiferromagnetic ordering temperature ($T{_N}$), leaving the Griffiths Temperature ($T{_G}$) invariant.  
\end{abstract}
\pacs{Pacs}
\maketitle
A Griffiths Phase (GP) pertains to the formation of arbitrarily large magnetically ordered regions \emph{within} the global paramagnetic phase, at temperatures exceeding that of long range magnetic ordering. First postulated in the context of random site dilution in Ising ferromagnets \cite{grif}, this phase is characterized by the magnetization remaining non-analytical at temperatures above the ferromagnetic Curie temperature ($T{_C}$), where the system is neither a pure paramagnet, nor does  it exhibit long range order. Analyticity is restored only above the Griffiths temperature $T{_G}$, which refers to the Curie Temperature of the pristine (undiluted) system. This model was further generalized to account for random magnetic spin systems, where $T{_G}$ now refers to the highest ordering temperature allowed by the bond probability distribution \cite{bray1}. The physics of this regime between $T{_C}$ and $T{_G}$ is now known to be extremely rich, and has even been extended to Quantum Phase Transitions where these effects are thought to manifest themselves in the form of power laws in thermodynamical observables \cite{sara}. This area of research saw renewed interest, when it was suggested that the phenomenon of Colossal Magnetoresistance observed in mixed valent manganites could be understood in the context of a Griffiths Singularity \cite{salamon}. A variety of factors like doping, Jahn Teller distortion \cite{JT}, size variance of the $A$ site ions in the $A$BO${_3}$ structure \cite{variance}, finite size effects \cite{size} and magnetic site dilution \cite{ashim} were seen to stabilize a GP in these materials. This phenomena appears to be ubiquitous to a number of strongly correlated electron systems, and materials as diverse as transition metal oxides \cite{JT, variance, size, ashim, cobaltate}, chemically substituted $f$-electron systems \cite{neto} and magnetocaloric intermetallics \cite{MCE0,MCE1} are now reported to harbor such a state.  

The probability of a magnetically ordered rare region existing within a global paramagnetic phase would fall exponentially as the volume of this region \cite{vojta} . Thus the experimental verification of the GP primarily relies on our ability to detect the contribution which this rare region makes to measurable thermodynamic quantities. The most popular route has been to evaluate the inverse of the measured zero field cooled (ZFC) dc magnetic susceptibility $\chi{^{-1}}=(T - T{_C})^{{1-\lambda}}$, where the non-universal exponent $\lambda$ exhibits positive values less than unity just above the magnetic ordering temperature \cite{salamon}. Within the GP, this measured magnetic susceptibility is the sum of the true paramagnetic (PM) susceptibility ($\chi{_{PM}}$) and the susceptibility of the magnetically ordered rare regions ($\chi{_{RR}}$). When the rare regions are ferromagnetic (FM), $\chi{_{RR}} > \chi{_{PM}}$, and a sharp downturn in $\chi{^{-1}}(T)$ is observed as one approaches $T{_C}$ from the high temperature side. On the contrary, if these rare regions are antiferromagnetic (AFM), this condition of $\chi{_{RR}} > \chi{_{PM}}$ may not be satisfied. Hence, it is not surprising that experimental reports of GPs have been scarce in systems exhibiting a PM to AFM  phase transition. The few examples which exist are systems which either have mixed interactions. or have a FM ground state in close proximity to the AFM one. For instance, in the inherently AFM half doped manganites, a GP was reported to be stabilized by the presence of short range FM superexchange interactions driven by structural distortions \cite{giri}. In the spin chain compound Ca${_3}$CoMnO${_6}$ with $\uparrow\uparrow\downarrow\downarrow$-type of magnetic ordering, competing AFM and FM interactions are known to result in a GP \cite{rao}. In the magnetocaloric intermetallics of the form $R{_5}$Si${_x}$Ge${_{1-x}}$(with $R$ = Gd, Tb, Dy, or Ho) with competing intralayer and interlayer interactions, local disorder is thought to promote the stabilization of FM short range correlations above the ordering temperature \cite{MCE1}. Very recently, a geometrically frustrated intermetallic system was reported to exhibit a GP in the vicinity of its PM - AFM transition \cite{ghosh}. However, this phase was also reported to originate from \emph{ferromagnetic} clusters arising as a consequence of its inherent non-stoichiometry. The only notable exception is the site diluted antiferromagnet Fe${_{1-x}}$Zn${_x}$F${_2}$, where the presence of a GP was proposed using a linear regression analysis of ac susceptibility data \cite{kleemann}. 

Antiferromagnetic DyBaCo$_4$O$_{7+\delta}$, the system under investigation here,  belongs to the class of geometrically frustrated antiferromagnets called the \emph{Swedenborgites}. This class of systems (of the form $A$Ba$Co{_4}O{_7}$, where typically $A=$Y, or a trivalent rare earth) have attracted attention due to the fact that the magnetic species (Co in this case) is exclusively tetrahedrally co-ordinated \cite{swede1}. The structure comprises of alternating triangular and Kagome slabs stacked along the crystallographic $c$ axis, and the frustration mandated by the lattice is typically relieved by a temperature driven structural phase transition which drives the system from a high symmetry (Trigonal $P31c$ or Hexagonal $P63mc$) phase to a low symmetry orthorhombic (typically $Pbn2_{1}$ or $Pna2_{1}$) phase \cite{swede2,panja, intTmCo} .We observe that with subtle variation of the synthesis conditions, polycrystalline specimens of DyBaCo$_4$O$_{7+\delta}$ can be \emph{preferentially} synthesized in either the high symmetry (HS) Trigonal $P31c$ or the low symmetry (LS) orthorhombic $Pna2_{1}$ phase, hitherto referred to as HS- or LS-DBCO respectively (Fig.S1 and Fig.S2 of Supplemental Material \cite{supple} ). The influence of the oxygen non-stoichiometry $\delta$ on the crystal structure of the Swedenborgites is known \cite{huq, avci}, and our HS and LS -DBCO specimens correspond to $\delta$ values of 0.13 and 0.07 respectively. 
\begin{figure}
	\centering
	\includegraphics[scale=0.23]{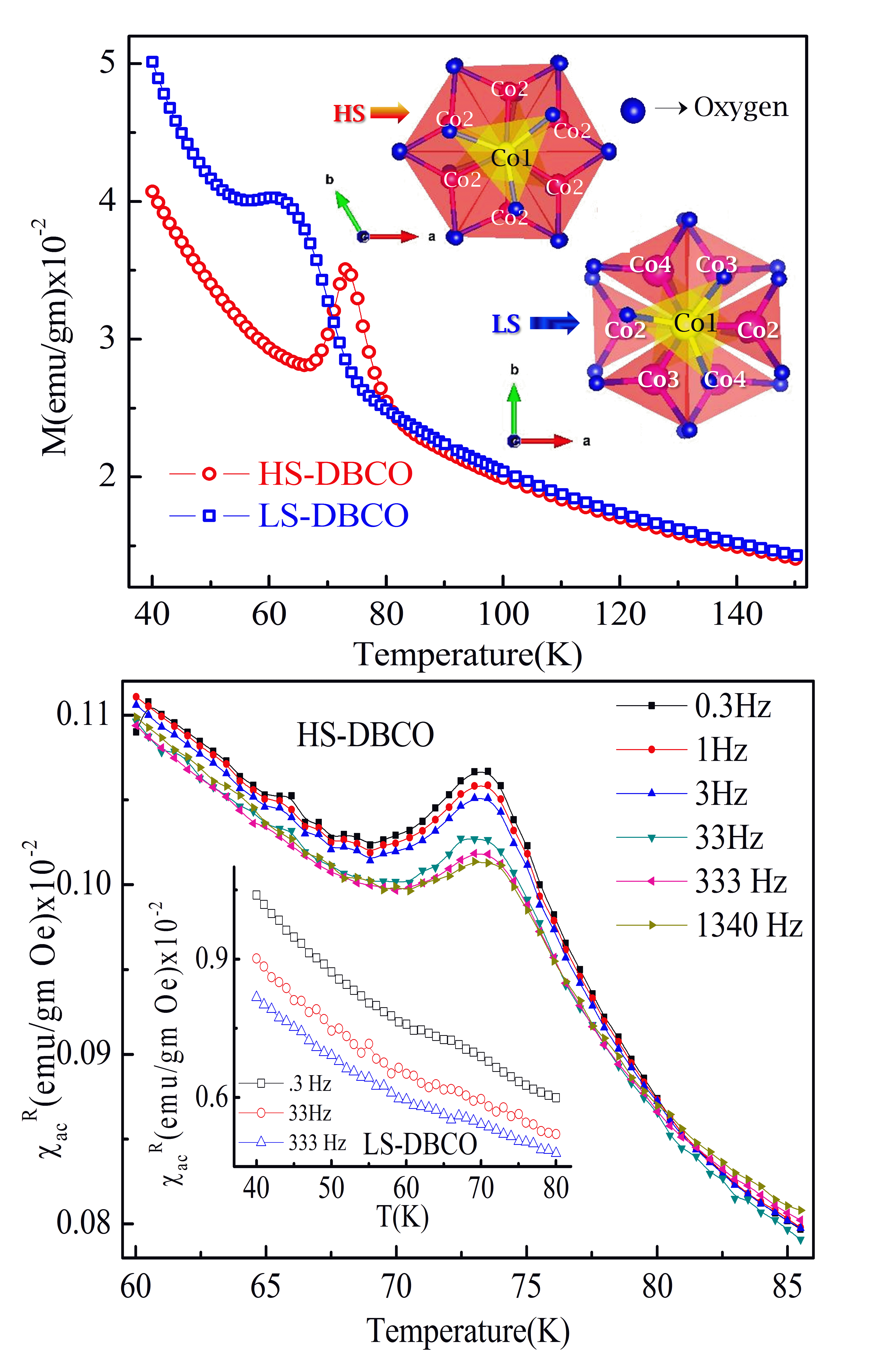}
	\caption{The zero field cooled dc magnetization  of the High Symmetry  and Low Symmetry DyBaCo$_4$O$_{7+\delta}$ specimens (top).  The distortion within the Kagome layer which distinguishes between the two structures is also depicted. The bottom panel exhibits  frequency dependent ac susceptibility measurements on both these specimens, ruling out the possibility of a glass-like state.}
	\label{Fig1}
\end{figure}
As shown in Fig .\ref{Fig1}, this transition from the high symmetry to low symmetry phase is characterized by a distortion of the perfect Kagome layer to one which comprises of three corrugated CoO${_4}$ tetrahedra. The LS-DBCO specimen is seen to exhibit a structural transition from a low symmetry $Pna2_{1}$ $\rightarrow$ high symmetry $P31c$ transition at temperatures in excess of 125 $\degree$ C (Fig.S3 of Supplemental Material \cite{supple} ). With Dy${^{3+}}$ having a spin only moment of 10.6 $\mu{_{\beta}}$, the magnetism in DyBaCo$_4$O$_{7+\delta}$ is plagued by the large paramagnetic contribution of Dy${^{3+}}$ ions. However, the onset of magnetic order within the Co sublattice in both the specimens can be clearly identified from conventional ZFC  magnetization measurements using a MPMS Squid magnetometer (Fig.\ref{Fig1} ), and the antiferromagnetic transition temperature ($T{_N}$) for the HS-DBCO and LS-DBCO specimens are seen to be 73K and 63K respectively. The lower panel shows temperature dependent measurements of the in-phase component of ac susceptibility $\chi{^R}{_{ac}}(T)$ for both the HS- as well as the LS-DBCO specimens, where the absence of a frequency dependent transition rules out the presence of a glass like phase in either of these systems. 

Measurements of the DC magnetic susceptibility as a function of temperature at different applied fields reveal telltale signatures of a GP in the LS-DBCO specimen.
\begin{figure}
	\centering
	\hspace{-0.32cm}
	\includegraphics[scale=0.23]{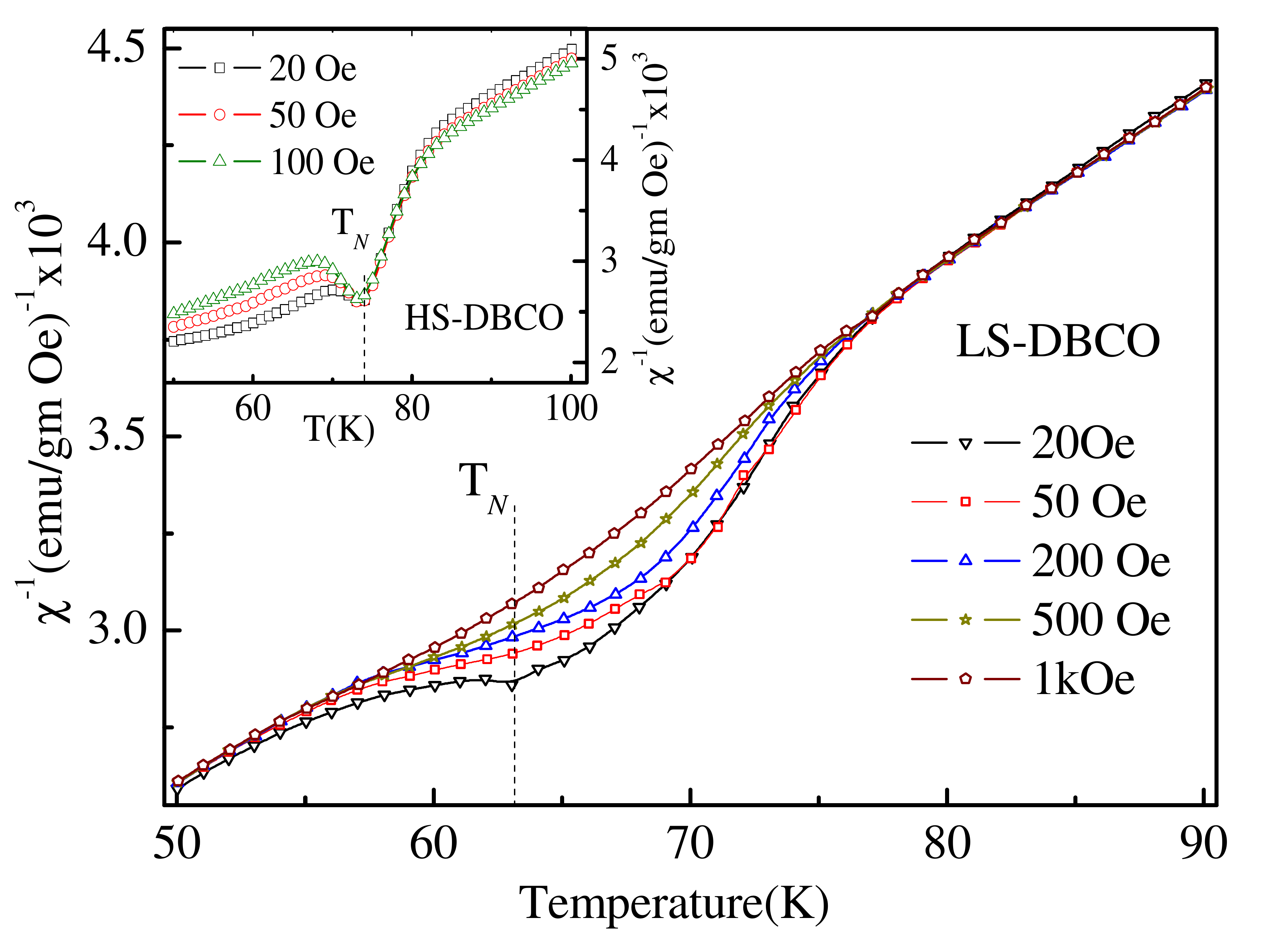}
	\caption{The inverse of the dc magnetic susceptibility as measured at different applied fields indicating the presence of a Griffiths Phase in the low-symmetry  DyBaCo$_4$O$_{7+\delta}$ .  The inset shows the absence of such a phase in the high symmetry specimen.}
	\label{Fig2}
\end{figure}
This is shown in the main panel of Fig.\ref{Fig2} where the inverse of the magnetic susceptibility ($\chi{^{-1}}$) exhibits a pronounced field dependent downturn on approaching the magnetic phase transition from the high temperature paramagnetic region. As described earlier, this downturn is expected when the susceptibility of the magnetically ordered rare regions is larger than that of the paramagnetic matrix, and application of moderate fields of the order of 1 kOe is sufficient to negate this condition in LS-DBCO. The onset of the GP can be deduced to be from $T{_G} = $76 K, which is very close to the $T{_N}$ of the high symmetry specimen. Interestingly, a similar measurement protocol does not indicate the presence of a GP in HS-DBCO, as is evident from the inset of Fig. \ref{Fig2} .\\ 

It is to be noted that for well after Griffiths theoretical work, the possibility of experimentally observing a Griffiths singularity was thought to be remote \cite{imry} . Though slow spin dynamics within the paramagnetic phase of the hole doped manganites was inferred from  muon spin scattering \cite{muon} and noise measurements \cite{noise}, these spatio-temporal inhomogenities were not directly connected to the possibility of a GP in these materials. A paradigm shift in experimental investigations of the GP in magnetic systems was the use of magnetic susceptibility measurements to infer on the possibility of a GP in the La${_{0.7}}$Ca${_{0.3}}$MnO${_3}$ system \cite{salamon}. Though subsequent work have used techniques like electron spin resonance \cite{JT}, specific heat \cite{sh}, and nonlinear magnetic susceptibility measurements \cite{ashim} to infer on the existence of a GP, in-field dc magnetic susceptibility measurements continue to be the technique of choice. As described earlier, though this appears to be sufficient in the case of ferromagnetic clusters, its efficacy in the investigation of a GP in a PM-AFM transition is suspect, since the susceptibility of the paramagnetic matrix ($\chi{_{PM}}$) is likely to be of the same order (or even exceed) the susceptibility of the magnetically ordered rare regions ($\chi{_{RR}}$). 

To circumvent this problem, we re-investigate the GP in the LS-DBCO specimen using the technique of Thermo-remanent magnetization (TRM). This measurement protocol has been used earlier in the investigation of spin glasses \cite{trm}, and involves cooling the specimen from well above its magnetic transition in the presence of an applied magnetic field, which is then switched off to zero at $T < T{_{C}}$. The magnetization is measured on warming up in this \emph{zero field} condition, and the thermo-remanent magnetization $M{_{TRM}}(T)$ typically exhibits a sharp upturn at the transition temperature. This technique has also found limited utility in the investigation of dilute antiferromagnets, and in the investigation of phenomena like piezomagnetism \cite{piezo1, piezo2, piezo3}. Being a zero-field measurement, it is likely to be advantageous in investigating the GP since the contribution of the paramagnetic susceptibility to the magnetization is likely to be suppressed in comparison with conventional in-field magnetic measurements. The upper panel of Fig. 3 exhibits $M{_{TRM}}(T)$ as measured in the LS-DBCO at different cooling fields. As is evident, a clear signature of the GP is seen in the form of an upturn of susceptibility at temperatures well above the $T{_N}$ of 63 K. The $T{_{G}}$ as inferred from this $M{_{TRM}}(T)$ measurement is 101 K - well in excess of that inferred from more traditional zero field cooled measurements depicted in Fig. \ref{Fig2}  reaffirming our contention that in-field magnetic measurements tends to underestimate the true $T{_{G}}$. 
\begin{figure}
		\centering
		\hspace{-0.85cm}
	\includegraphics[scale=0.25]{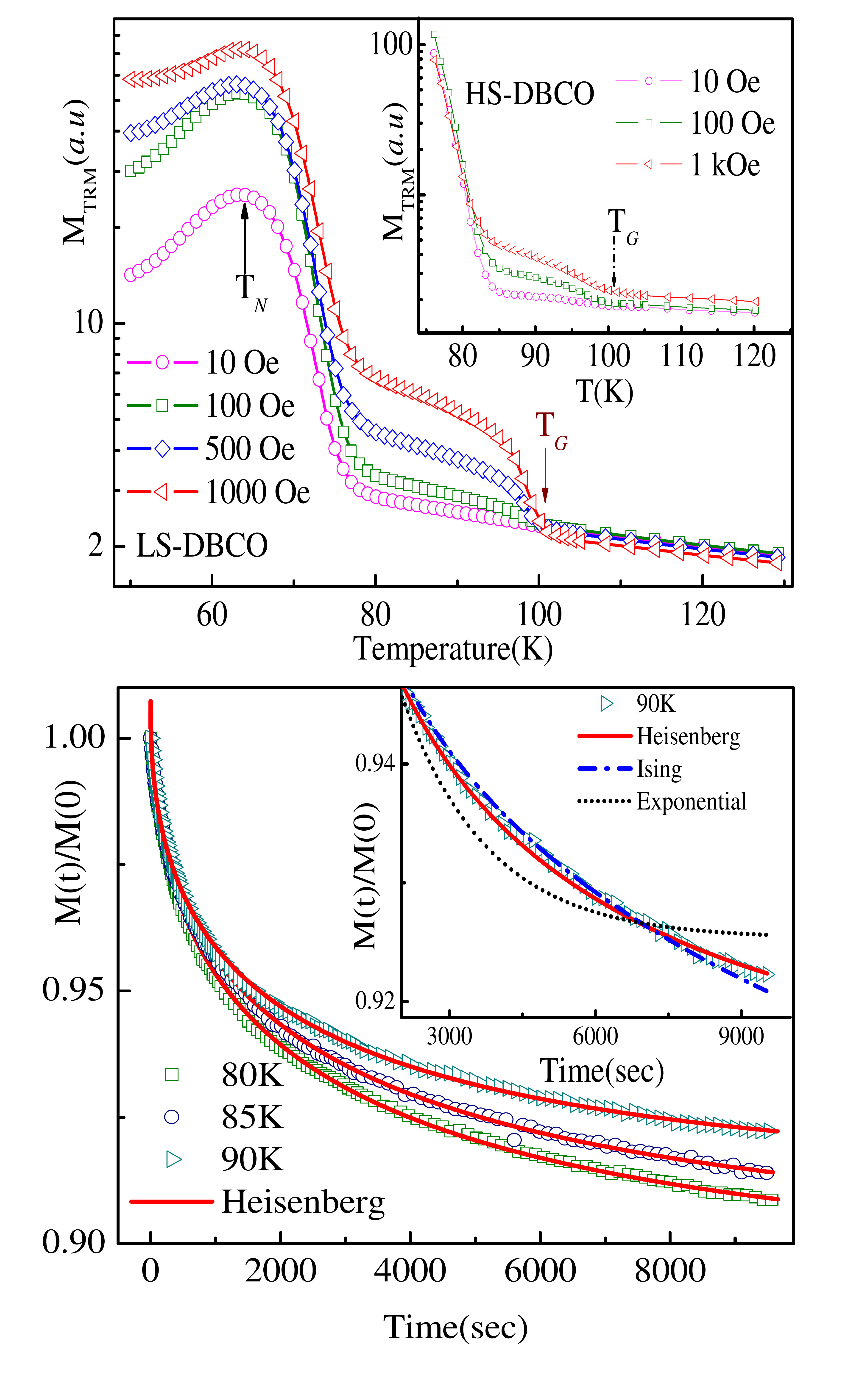}
	\caption{Measurements of the thermoremanent magnetization ($M{_{TRM}}$) as measured for the Low Symmetry  DyBaCo$_4$O$_{7+\delta}$ specimen at different cooling fields, showing the onset of the Griffiths Phase ($T{_{G}}$), and the antiferromagnetic transition temperature ($T{_N}$). The inset depicts the onset of the Griffiths Phase ($T{_{G}}$) in the High Symmetry DyBaCo$_4$O$_{7+\delta}$ specimen. The lower panel shows the time decay of the Isothermal remanent magnetization at $T{_N}<T<T{_G}$ for the LS-DBCO specimen, fitted with a $exp(-Bt^{0.5})$ form expected for Heisenberg spin systems. The inset  also shows the relative lack of fitting with an exponential decay and with the form expected for an Ising system.}
	\label{Fig3}
\end{figure}

The presence of magnetically ordered rare regions within the GP is expected to drastically influence the dynamics of the magnetic susceptibility. In the high temperature paramagnetic limit, the susceptibility would be expected to fall exponentially as a function of time. However, in the Griffiths phase, the time required to reverse the effective spin of large magnetically ordered clusters would be larger, thus resulting in an non-exponential decay of the spin-autocorrelation function \cite{mohit}. For randomly diluted  magnets, this decay has been calculated to be of the form $exp[-A(ln t^{d/{d-1}}]$ and $exp(-Bt^{0.5})$ for Ising and Heisenberg spin systems respectively \cite{bray}  . This slowing down would be expected to manifest itself in the form of a power law in measurements of the magnetization decay measured as a function of time. The lower panel of Fig.\ref{Fig3} shows measurements of the isothermal remanent magnetization (IRM), for the LS-DBCO specimen within the GP regime. These measurements were performed by cooling the system from 300 K at an applied magnetic field of 500 Oe, and measuring the decay of the magnetization as a function of time on switching off the magnetic field. Our measurements indicate that the observed decay follows a functional form expected for systems where the interaction between spins are Heisenberg like. The lack of agreement with the functional form expected for Ising-like interactions, and an exponential decay is also shown in the inset. Our observations are in broad agreement with prior data, where the nature of spin correlations in the Cobalt based Swedenborgites was inferred to be Heisenberg-like from neutron scattering experiments \cite{pascal}. 

Considering the utility of TRM measurements in identifying the presence of a GP, we re-investigate the HS-DBCO specimen using this measurement protocol. Interestingly, our $M{_{TRM}}(T)$ measurements indicate that a GP is \emph{also} present in the HS-DBCO as is shown in the inset of the upper panel in Fig.\ref{Fig3}  . 
\begin{figure}
	\centering
	\hspace{-0.3cm}
	\includegraphics[scale=0.24]{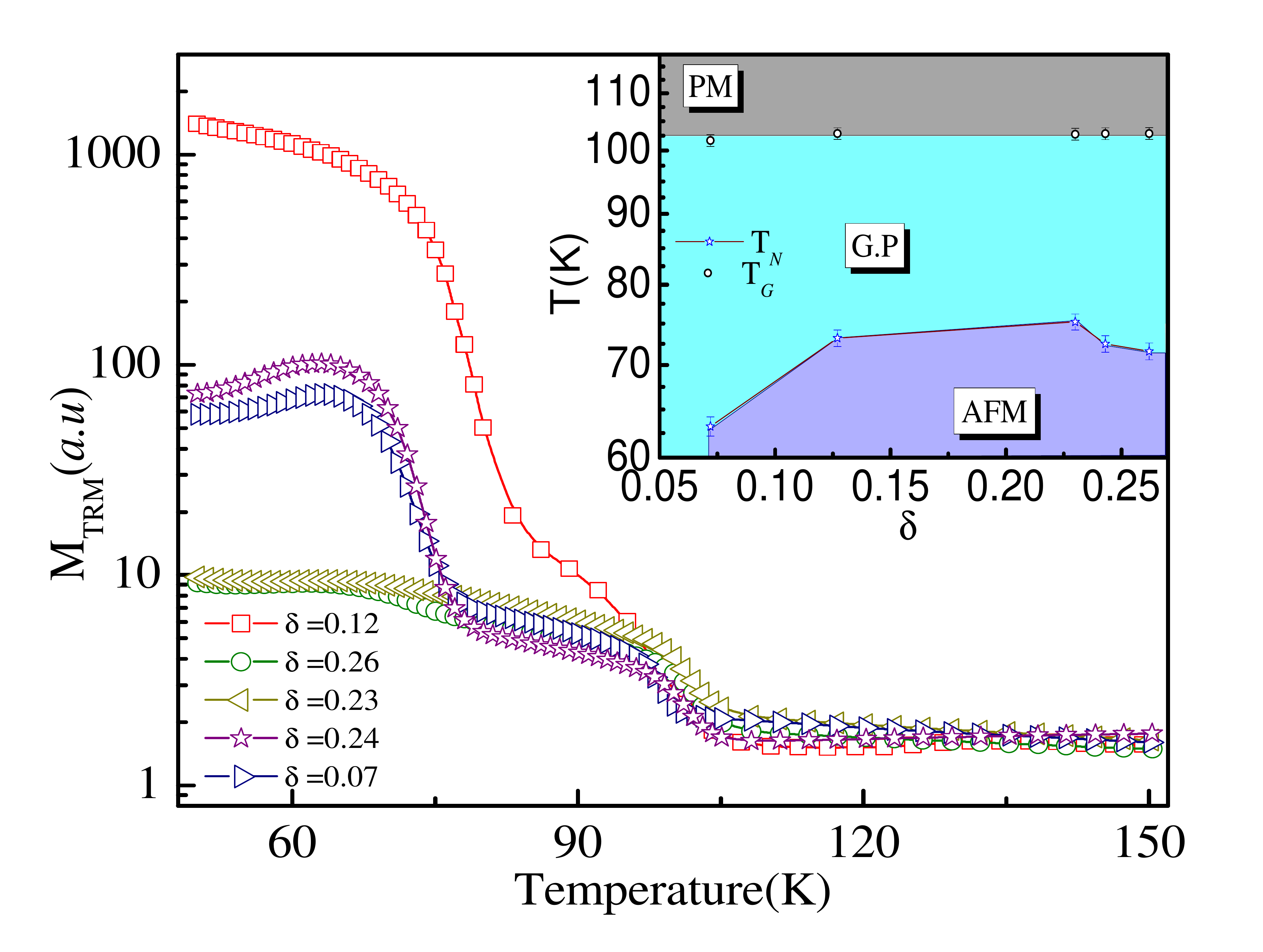}
	\caption{ Measurements of the thermoremanent magnetization ($M{_{TRM}}$) as measured for different  DyBaCo$_4$O$_{7+\delta}$ specimens at a cooling field of 1 KOe. The inset shows the  relative invariance of the onset of the Griffiths Phase for specimens with different Oxygen nonstoichiometry $\delta $ .  }
	\label{Fig4}
\end{figure}
The fact that in-field measurements exhibited no traces of such a phase demonstrates the superiority of the TRM protocol over conventional ZFC measurements in the identification of the GP. Moreover, the $T{_G}$ is also seen to be invariant of the room temperature crystallographic symmetry, and both specimens exhibit similar values of $T{_G} \approx$ 101 K. To further corroborate this observation, we have also synthesized a number of specimens with varying levels of oxygen non-stoichiometry $\delta$ with slight modifications of the synthesis conditions. As can be seen from Fig.\ref{Fig4}, though the onset of long range magnetic order varies across the different specimens, the onset of the GP is invariant, with all specimens exhibiting a $T{_G}$ $\approx$ 101 K. Thus  $T{_G}$ appears to be relatively insensitive to changes in the Co${^{2+}}$/Co${^{3+}}$ ratio, as well as to subtle variations in the crystallographic symmetry. 

Systems stabilizing in geometrically frustrated motifs are known to have an extended region in phase space, where short range correlations are prevalent. In the Swedenborgites, the presence of trigonal bipyramid clusters satisfying the $\displaystyle \sum_{i} S{_i} = 0$ condition have been inferred from prior neutron scattering experiments in the closely related YBaCo${_4}$O${_7}$ system \cite{pascal}  . Monte Carlo simulations  have indicated that the nature of spin correlations can be understood by modeling the parameter $J_{out}/J{_{in}}$, where $J{_{out}}$ and $J{_{in}}$ refer to the exchange interaction strength between the Kagome and the triangular layers, and within the Kagome layer respectively \cite{khalyavin}. The magnetic diffuse scattering observed above the ordering temperature in YBaCo${_4}$O${_7}$ could be modeled by using  $J{_{out}}/J{_{in}}$ $\approx 1$, which corresponded to quasi-1D magnetic correlations along the $c$ axis. Other members of the Swedenborgite family would correspond to different values of $J{_{out}}/J{_{in}}$, with this ratio being determined by factors such as the ionic radii of the $A$ site cation, and the preferential occupation of the Kagome and triangular sublattices by Co${^{2+}}$ and Co${^{3+}}$ ions. The specific experimental signatures evident in our bulk magnetic measurements indicate that the \emph{onset} of these short range correlations could be looked upon as a Griffiths Singularity. The fact that $T{_G}$ is invariant of both the room temperature structure, and the oxygen non-stoichiometry $\delta$ indicate that these factors do not appear to drastically modulate the effective exchange interactions

In summary, we report on the identification of a GP in a geometrically frustrated antiferromagnet DyBaCo$_4$O$_{7+\delta}$ using bulk dc magnetic measurements. Conventional in-field measurements tend to underestimate the true $T{_G}$, and measurements using the TRM protocol outlined here is seen to offer an effective means of identifying such transitions. The temporal evolution of magnetization is seen to follow a functional form expected for Heisenberg spin systems. Measurements on specimens with varying oxygen stoichiometry $\delta$ indicate that the onset of this GP ($T{_G}$) is relatively insensitive to variables such as the Co${^{2+}}$/Co${^{3+}}$ ratio. The measurement protocol described here would find utility in the detection of GPs in a number of strongly correlated materials. We also expect this work to spur further experimental and theoretical investigations into the presence of Griffith singularities in geometrically frustrated magnets.    

The authors acknowledge Nilesh Dumbre for technical assistance in x-ray diffraction measurements. S.N. acknowledges DST India for support through grant no. SB/S2/CMP-048/2013.
\bibliography{Bibliography_mod}
\end{document}